\documentclass[12pt,aps,floats,amsmath,showpacs]{revtex4}
\input BoxedEPS
\SetRokickiEPSFSpecial  %% for dvips by Tom Rokicki, for VMS
\HideDisplacementBoxes

\newcommand{\Cite}[1]{[{\tt #1}]\cite{#1}\relax}
\let\Cite\cite

\newcommand{\numeq}[2]{\begin{equation}
#2
\label{#1}
\end{equation}}

\newcommand{\bra}[1]{\bigl\langle #1 \bigr| }
\newcommand{\ket}[1]{\bigl| #1 \bigr\rangle}
\newcommand{\q}{\quad}

\begin{document}
%\draft
%\tighten
 \title{Effect of recent \protect\boldmath$R_p$ and $R_n$ measurements\\ on
extended Gari-Kr\"umpelmann model\\ fits  to nucleon  electromagnetic form
factors}

% repeat the \author\address pair as needed

\author{Earle L. Lomon}
\affiliation{Center for Theoretical Physics\\
Laboratory for Nuclear Science and 
Department of Physics\\ Massachusetts Institute of Technology, Cambridge, 
Massachusetts 02139\\
\rm MIT-CTP-3257}
%\date{}

\begin{abstract}\noindent
% insert abstract here. 
The Gari-Kr\"umpelmann (GK) models of nucleon electromagnetic form factors, in which
the $\rho$, $\omega$, and  $\phi$ vector meson pole contributions evolve at  high momentum
transfer to conform to the predictions of perturbative QCD (pQCD),  was recently extended to
include the width of the $\rho$ meson by substituting the result of dispersion relations for
the pole and the addition of the $\rho'$ (1450) isovector vector meson pole.  This extended model
was shown to produce a good overall fit to all the available nucleon electromagnetic form
factor (emff) data.  Since then new polarization data shows that the electric to magnetic ratios 
$R_p$ and $R_n$ obtained are not consistent with the older $G_{Ep}$ and $G_{En}$ data in
their range of momentum transfer. The model is further extended to include the $\omega'$
(1419) isoscalar vector meson pole.  It is found that while this GKex cannot simultaneously fit
the new $R_p$ and the old $G_{En}$ data, it can fit the new $R_p$ and $R_n$ well
simultaneously.  An excellent fit to all the remaining data is obtained when the inconsistent
$G_{Ep}$ and $G_{En}$ is omitted.  The model predictions are extended beyond the data,
if needed, to momentum transfer squared, $Q^2$, of 8  GeV$^2/c^2$.
\end{abstract}


\maketitle

% insert suggested PACS numbers in braces on next line

% body of paper here
\section{INTRODUCTION}

	A variety of related models of the nucleon emff \Cite{EL} were fitted to the
complete set of data available before September 2001.  One group of models
included variants of the basic GK model of $\rho$,
$\omega$, and  $\phi$ vector meson pole terms with hadronic form factors and a term with
pQCD behavior which dominates at high $Q^2$ \Cite{GK}.  Four varieties of hadronic form factor
parameterization (of which two are used in \Cite{GK}) were compared.   In addition to the GK
type models we considered a group of models (generically designated DR-GK) that use the
analytic approximation of \Cite{MMD} to the dispersion integral  approximation for the $\rho$
meson contribution (similar to that  of \Cite{HO}), modified by the four hadronic form  factor
choices used with the GK model, and the addition of the well established $\rho'$ (1450) pole. 
Every model had an electric and a magnetic coupling parameter for each of the three pole
terms, four ``cut-off" masses for the hadronic form-factors and the QCD scale mass scale,
$\Lambda_{\mathrm{QCD}}$ for the logarithmic momentum transfer behavior in pQCD.  In
addition the effect of a normalization parameter was sometimes considered for the dispersion
relation behavior of the $\rho$ meson in the DR-GK models.

	When the set of parameters in each of the eight models was fitted to the full set of data
available before publication,  for $G_{Ep}$, $G_{Mp}$, $G_{En}$, $G_{Mn}$ and  the lower $Q^2$
values of $R_p \equiv \mu_p G_{Ep}/G_{Mp}$, three GK and all four DR-GK models attained
reasonable $\chi^2$ (when the inconsistency of some low $Q^2$ $G_{En}$ and $G_{Mn}$ data
was taken into account), but the extended DR-GK models had significantly lower 
$\chi^2$.  Furthermore $\Lambda_{\mathrm{QCD}}$ was reasonable for three of the DR-GK
models but for only the one of the GK models that had an unreasonably large  anomalous
magnetic coupling $\kappa_\rho$.  It was concluded that the three DR-GK models were the best
nucleon emff to use in prediction of nuclear electromagnetic properties.  All thee were found
to be moderately consistent in their predictions up to $Q^2$ of 8 ~GeV$^2/c^2$.

	However the part of the above data set from a recent $R_p$  ratio data \cite{CP1} for $0.5 
$~GeV$^2/c^2 \leq Q^2 \leq 3.5$ ~GeV$/
^2/c^2$, swamped statistically by all the other
data, was systematically lower than the fitted models (Fig.5 of \cite{EL}) contributing
disproportionately to $\chi^2$.  This ratio is determined by an asymmetry measurement in the
scattering of polarized electrons on protons.  Multiplied by the well determined values of $G_{Mp}$
one obtains values for $G_{Ep}$ which are not subject to the uncertainty inherent in the Rosenbluth
separation measurements in which $G_{Ep}$ is obtained by subtracting the much larger
contribution of $G_{Mp}$ from the unpolarized cross section.  As expected the $G_{Ep}$ derived
from the measured $R_p$ are consistently below those of the older Rosenbluth separation values.
	
\begin{figure}[htbp]
$$
\BoxedEPSF{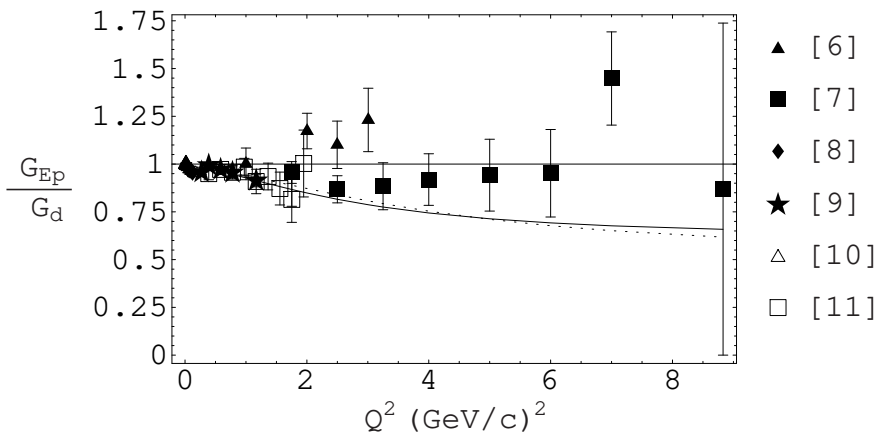 scaled 1200}
$$
\caption{$G_{Ep}$ normalized to $G_d$, comparing the GKex(01) fit [dotted] with the fit GKex(01-)
[solid] obtained when the data of \protect\Cite{RCW} is omitted.  The other $ G_{Ep}$ data is from 
\cite{LA}, \cite{FB}, \cite{KMH}, \cite{JJM}, and \cite{CHB}  The data symbols are
listed in the figure.}
\label{elFig1}
\end{figure}

\begin{figure}[hbt]
$$
\BoxedEPSF{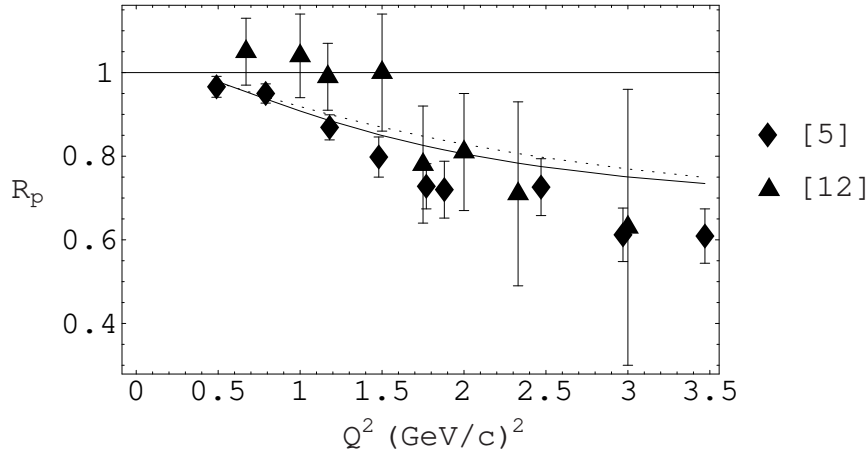 scaled 1200}
$$
\caption{$R_p$, the ratio $\mu_p G_{Ep}/G_{Mp}$,  comparing the GKex(01) fit [dotted] with the fit 
GKex(01-) [solid] obtained when the data of \protect\Cite{RCW} is omitted.  The data is from 
\protect\Cite{CP1} and \cite{WB73}.  The data symbols are listed in the figure.}
\label{elFig2}
\end{figure}

	It is plausible to expect that the old $G_{Ep}$ data is responsible for restricting the best
 fit of the
models  to be substantially above the experimental $R_p$ values.  With this in mind the
particularly high data of \cite{RCW} was omitted from the fit to the model type DR-GK$'$(1) of
\cite{EL} and the flexibility of a $\rho$ meson dispersion integral normalization parameter N was
included.  In this article the original version is designated as GKex(01) and when fitted to the
smaller data set as GKex(01-).  As seen in Tables~\ref{T1} and~\ref{T2}
 and Figs.~\ref{elFig1} and
\ref{elFig2}, there is only a small change in the fit to $G_{Ep}$ and $R_p$, although the parameters 
of the fit change substantially.

	After the publication of \cite{EL} new data \cite{CP2} extended the measurements of 
$R_p$ up to
$Q^2 = 5.6$ ~GeV$^2/c^2$, exacerbating the discrepancy with the predictions of the best models in
\cite{EL}.  Very recently $R_n \equiv \mu_n G_{En}/G_{Mn}$ has been obtained directly
\cite{RM} by the scattering of polarized electrons on deuterium and detecting the polarized recoil
neutron at $Q^2$ = 0.45, 1.15 and 1.47  ~GeV$^2/c^2$.  The preliminary results are consistent with
the Galster \cite{Ga} parameterization from lower $Q^2$ data
\numeq{e1} {
R^{Galster}_n(Q^2) = -\frac{\mu_n \tau}{1 + 5.6 \tau},\q \tau =
\frac{Q^2}{4m_N^2}\ .
}
which, in parallel to the situation for $R_p$, implies much lower values of $G_{En}$ in their $Q^2$
range when coupled with $G_{Mn}$ values (either the precision data of \cite{HA1} or the model fits).

	In this paper, in addition to the above comparison of GKex(01) and GKex(01-), we fit the model
of type DR-GK$'$(1), with the added isoscalar vector meson $\omega'$(1419) pole, to the following
data sets, chosen to determine the effect of the old $G_{En}$ and $G_{Ep}$ data in direct conflict
with the values of $R_n$ and $R_p$ from modern polarization measurements:
\begin{enumerate}
	\item[(a)]  The fit GKex(02L) from the full data set of \cite{EL} with the addition
of
\cite{CP2} and
\cite{RM}, the omission of \cite{RCW} (as above for GKex(01-)) and the $G_{En}$ values for 
$Q^2 \geq 0.779$ ~GeV$^2/c^2$ of \cite{KMH},  \cite{AL}, and \cite{WB72}. 
\item[(b)]  The fit of GKex(02S) to the same data set as above except for the
omission of the $G_{Ep}$ values for $Q^2 \geq 1.75$ ~GeV$^2/c^2$ of \cite{LA}.
\end{enumerate}
It will be seen that the omission of the conflicting $G_{En}$ data, GKex(02L), has a much bigger
influence than the omission of \cite{RCW}, GKex(01-), enabling a much better fit to $R_p$ in
addition to a very good fit to $R_n$, compared to GKex(01).  With the removal of
the conflicting
$G_{Ep}$ data, GKex(02S), the fit to all the remaining data, including $R_p$, is very satisfactory.

	In Section~\ref{s:2} we will specify the models and parameters used in this
 article,  and the data sets used in Section~\ref{s:3}.  In
Section~\ref{s:4} we present the results of the four GKex fits in comparison with each other.  We 
extrapolate beyond the present experimental range of momentum transfer where necessary for
predicting available deuteron emff data.  The model GKex(02S) fits the modern and consistent older 
data well
and meets the requirements of dispersion relations and of QCD at low and high
momentum transfer.  Conclusions are presented in Section~\ref{s:5}. 
\newpage

\section{The Nucleon emff model}\label{s:2}

	In fitting the nucleon emff data including the new $R_n$ and $R_p$ results we have chosen to use
the extended GK model DR-GK$'$(1) of Ref. \cite{EL} with the addition of a pole term for the well
established isoscalar vector meson $\omega'$(1419), whose mass is lower than that of the already
included isovector vector meson $\rho'$(1450).  We denote this model as GKex.  The choice of the
particular parameterization DR-GK$'$(1) was made because of its low $\chi^2$ value and the fact
that its predicted values of
$R_p$ were a little closer to the data than those of the other extended models.  In addition
DR-GK$'$(1) has the following good physical properties:
\begin{enumerate}
\item[(1)]  It uses the QCD cut-off $\Lambda_2$ for the helicity flip
meson-nucleon form factors, rather than the meson cut-off $\Lambda_1$ used
by DR-GK(3) and DR-GK$'$(3). 
\item[(2)]  The evolution of the logarithmic dependence on $Q^2$ is controlled by
the quark-nucleon cut-off $\Lambda_D$, along with
$\Lambda_{\mathrm{QCD}}$.  DR-GK(1) and DR-GK(3) use $\Lambda_2$
instead of $\Lambda_D$. 
\item[(3)]  Fitted to the data set of \cite{EL} it finds $\Lambda_{\mathrm{QCD}}$
= .1163, close to the expected value.  The form factors are not very sensitive to
this parameter which is fixed at .15  for the fits to the new data sets.
\end{enumerate}

	So that the reader need not make constant reference to \cite{EL} we repeat the relevant 
formulas
here together with the new $\omega'$(1419) terms.

	The emff of a nucleon are defined by the matrix elements of the 
electromagnetic current~$J_\mu$
\numeq{e2}{
\bra{N(p')}\, J_\mu \,\ket{N(p)} = e \bar u (p') \Bigl\{
\gamma_\mu F^N_1 (Q^2) + \frac i{2m_N} \sigma_{\mu\nu} Q^\nu F^N_2(Q^2)
\Bigr\} u(p) 
}
where $N$ is the neutron, $n$, or proton, $p$, and $-Q^2= (p'-p)^2$ is the square  of
the invariant momentum transfer.  $F_1^N(Q^2)$ and $F_2^N(Q^2)$ are respectively 
the  Dirac and Pauli form factors, normalized at $Q^2=0$ as
\numeq{e3}{
F^p_1(0) = 1,\q F^n_1(0)=0,\q F^p_2(0) = \kappa_p,\q F^n_2(0)= \kappa_n\ .
}
The Sachs form factors, most directly obtained from experiment, are then
\begin{eqnarray}
G_{\mathrm{EN}} (Q^2) &=& F^N_1 (Q^2) -\tau F^N_2(Q^2)\nonumber\\
G_{\mathrm{MN}} (Q^2) &=& F^N_1 (Q^2) +  F^N_2(Q^2)\ .
\label{e4}
\end{eqnarray}
Expressed in terms of the isoscalar and isovector electromagnetic currents
\numeq{e5}{
2F^p_i = F^{\null{is}}_i + F^{\null{iv}}_i, \q
2F^n_i = F^{\null{is}}_i - F^{\null{iv}}_i, \q
(i=1,2)\ .
}

The GKex model has the following form for the four isotopic emff:
\begin{align}
F^{\null{iv}}_1(Q^2) &= N/2
\frac{1.0317 + 0.0875(1+Q^2/0.3176)^{-2}}{(1+Q^2/0.5496)}
F^\rho_1(Q^2)\nonumber\\
&\qquad{}  + \frac{g_{\rho'}}{f_{\rho'}}
\frac{m^2_{\rho'}}{m^2_{\rho'} + Q^2} F^\rho_1(Q^2)
+ \Bigl( 1-1.1192\, N/2 -  \frac{g_{\rho'}}{f_{\rho'}} \Bigr) F^D_1(Q^2)
\nonumber\\
F^{\null{iv}}_2(Q^2) &=  N/2 \frac{5.7824 +
0.3907(1+Q^2/0.1422)^{-1}}{(1+Q^2/0.5362)}
F^\rho_2(Q^2)\nonumber\\
 &\qquad{}  + \kappa_{\rho'} \frac{g_{\rho'}}{f_{\rho'}}
\frac{m^2_{\rho'}}{m^2_{\rho'} + Q^2} F^\rho_2(Q^2)
+ \Bigl( \kappa_\nu -6.1731\,N/2 -  \kappa_{\rho'}\frac{g_{\rho'}}{f_{\rho'}} \Bigr)
F^D_2(Q^2)\nonumber\\
F^{\null{is}}_1(Q^2) &=  \frac{g_\omega}{f_\omega}
\frac{m^2_\omega}{m^2_\omega + Q^2} F^\omega_1(Q^2) +
\frac{g_{\omega'}}{f_{\omega'}}
\frac{m^2_{{\omega'}}}{m^2_{\omega'} + Q^2} F^\omega_1(Q^2) + 
\frac{g_\phi}{f_\phi}\frac{m^2_\phi}{m^2_\phi + Q^2} F^\phi_1(Q^2)\nonumber\\
&\qquad{} +\Bigl(
1-\frac{g_\omega}{f_\omega}-\frac{g_{\omega'}}{f_{\omega'}}\Bigr)
F^{\null D}_1(Q^2)\label{e6}\\
F^{\null{is}}_2(Q^2) &= \kappa_\omega \frac{g_\omega}{f_\omega}
\frac{m^2_\omega}{m^2_\omega + Q^2} F^\omega_2(Q^2) +
\kappa_{\omega'} \frac{g_{\omega'}}{f_{\omega'}}
\frac{m^2_{\omega'}}{m^2_{\omega'} + Q^2} F^\omega_2 (Q^2) + 
\kappa_\phi \frac{g_\phi}{f_\phi}\frac{m^2_\phi}{m^2_\phi + Q^2} F^\phi_2 (Q^2)\nonumber\\
&\qquad{} + \Bigl(
\kappa_{\null s} - \kappa_\omega \frac{g_\omega}{f_\omega} - \kappa_{\omega'} 
\frac{g_{\omega'}}{f_{\omega'}}-\kappa_\phi\frac{g_\phi}{f_\phi}\Bigr) F^{\null D}_2(Q^2) \nonumber
\end{align}

where the pole terms are those of the $\rho$, $\rho'$, $\omega$, $\omega'$, and  $\phi$  mesons,
and the  final term of each equation is determined by the asymptotic properties of
PQCD.   The $F_i^\alpha$, $\alpha = \rho$,  $\omega$, or $\phi$ are the
meson-nucleon form factors,  while the $F_i^D$ are effectively quark-nucleon form
factors.

For GKex the above  hadronic form factors are parameterized in the following way:
\begin{eqnarray}
F^{\alpha,D}_1(Q^2) &=&  \frac{\Lambda^2_{1,D}}{\Lambda^2_{1,D} +\tilde Q^2}
\frac{\Lambda^2_2}{\Lambda^2_2 +\tilde Q^2}\nonumber\\
F^{\alpha,D}_2(Q^2) &=&   \frac{\Lambda^2_{1,D}}{\Lambda^2_{1,D} +\tilde
Q^2}
\Bigl(\frac{\Lambda^2_2}{\Lambda^2_2 +\tilde Q^2}\Bigr)^2\nonumber\\[1ex]
\noalign{\mbox{where $\alpha=\rho,\omega$ and $\Lambda_{1,D}$ is $\Lambda_{1}$ for
$ F^\alpha_i$, $\Lambda_D$ for $ F^D_i$,}\medskip}   %\nonumber\\
F^\phi_1(Q^2) &=& F^\alpha_1\Bigl(\frac{Q^2}{\Lambda^2_1 + Q^2}
\Bigr)^{1.5}\ , \quad F^\phi_1(0) = 0\nonumber\\
F^\phi_2(Q^2) &=&
F^\alpha_2\Bigl(\frac{\Lambda^2_1}{\mu^2_\phi}\frac{Q^2+\mu^2_\phi}{\Lambda^2_1
+ Q^2}
\Bigr)^{1.5}
\label{e7} \\
\mbox{with} \tilde Q^2 &=& Q^2\frac{\ln\bigl[(\Lambda^2_D +
Q^2)/\Lambda^2_{\mathrm{QCD}}\bigr]}{\ln(\Lambda^2_D/\Lambda^2_{\mathrm{QCD}})}\ . \nonumber
\end{eqnarray}
This parameterization, together with Eq.~(\ref{e6}), guarantees that the 
normalization conditions of Eq.~(\ref{e2}) are met and that asymptotically
\begin{align}
F^i_1 &\sim \bigl[ Q^2 \ln(Q^2/\Lambda^2_{\mathrm{QCD}})\bigr]^{-2}\nonumber\\
F^i_2 &\sim F^i_1/Q^2 \label{e8}\\
i &= is, iv \nonumber
\end{align}
as required by PQCD.  The form factor $F^\phi_1(Q^2)$ vanishes at  $Q^2 =0$, and it and 
$F^\phi_2(Q^2)$ decrease more rapidly at large  $Q^2$ than the other meson form factors.  This
conforms to the Zweig rule imposed by the $s\bar s$ structure of the $\phi$ meson \cite{GK}.

	This model has at most 14 free parameters:
\begin{enumerate}
\item[(i)]   8 couplings to the pole terms,  the 4 $g_m/f_m$ and the 
     4 $\kappa_m$ for the $\rho'$, $\omega$, $\omega'$ and $\phi$ mesons.
\item[(ii)]  4 cut-off masses in the hadronic form factors, $\Lambda_1$, 
     $\Lambda_2$, $\Lambda_D$,  and $\mu_\phi$.
\item[iii)] The mass determing the size of the logarithmic $Q^2$ behavior,
$\Lambda_{\mathrm{QCD}}$. 
\item[iv)] The normalization factor N for the dispersion
relation contribution of the $\rho$ meson.
\end{enumerate}

However at most 12 of these
parameters are freely varied in any of the fits made in the next section to the
chosen data sets.

\section{Data base and fitting procedure}\label{s:3}

	As previously stated, GKex(01) is the same as DR-GK$'$(1) of \cite{EL}.  
This model had the best fit to the full
data set available at the publication of \cite{EL} with $g_\omega'/f_\omega'$ = 
$\kappa_\omega'$ = 0 and with N=1.  For GKex(01-) the 4 data points of \cite{RCW} were omitted
from that data set.   In this case $g_\omega'/f_\omega'$ and $\kappa_\omega'$ were still
supressed but N was freely varied.

	In the fits GKex(02L) and GKex(02S) $g_\omega'/f_\omega'$ and $\kappa_\omega'$ were freely
varied, but these fits fixed N=1 again (implying negligible error in the dispersion relation 
evaluation) and
$\Lambda_{\mathrm{QCD}}$ was fixed at the physical value of 0.15 ~GeV$/c$.  The important
difference from the data set of GKex(01-) is the addition of the higher Q  $R_p$ data points of
\cite{CP2} and the  $R_n$ data points of \cite{RM} and the omission of the $G_{En}$ values for 
$Q^2 \geq 0.779$ ~GeV$^2/c^2$ of \cite{KMH}, \cite{WB72} and \cite{AL}.  In the shorter data set of
GKex(02S) the $G_{Ep}$ values for $Q^2 \geq 1.75$ ~GeV$^2/c^2$ of \cite{LA} are also omitted. 
The free parameters were optimized using a Mathematica program that incorporates the
Levenberg-Marquardt  method.

\begin{table}[p]
 \caption{Model parameters.  Common to all models are $\kappa_v=3.706$,
$\kappa_s=-0.12$, $m_\rho=0.776$ GeV, $m_\omega=0.784$ GeV, $m_\phi=1.019$
GeV,  $m_{\rho'}= 1.45$ GeV and $m_{\omega'}= 1.419$ GeV.}\medskip
 \label{T1}
 \begin{tabular}{c|cccc}
Parameters & & Models &  \\
\hline\hline
 &  GKex(01)&  GKex(01-) & GKex(02L) & GKex(02S) \\
$g_(\rho')/f_(\rho')$ & 0.0636 & 0.0598 & 0.0608 & 0.0401 \\
$\kappa_(\rho')$ & $-0.4175$ & $-15.9227$ & 5.3038 & 6.8190 \\
$g_\omega/f_\omega$ & 0.7918 & 0.6981 & 0.6896 & 0.6739 \\ $\kappa_\omega$ & 5.1109 &
1.9333 & $-2.8585$ & 0.8762 \\
$g_\phi/f_\phi$ & $-0.3011$ & $-0.5270$ & $-0.1852$ & $-0.1676\ \  $ \\  $\kappa_\phi$ &
13.4385 & 2.3241 & 13.0037 &  7.0172
\\ $\mu_\phi$ & 1.1915 & 1.5113 & 0.6848 & 0.8544\\
$g_(\omega')/f_(\omega')$ & & & 0.2346 & 0.2552 \\
$\kappa_(\omega')$ & & & 18.2284 & 1.4916\\
$\Lambda_1$ & 0.9660 & 1.1276 & 0.9441 & 0.9407 \\ $\Lambda_D$ & 1.3406 &
1.8598 &  1.2350 & 1.2111  \\ $\Lambda_2$ & 2.1382 & 1.2255 & 2.8268 & 2.7891  \\
$\Lambda_{\mathrm{QCD}}$
& 0.1163 & 0.1315 & 0.150$^{\textrm{(a)}}$ & 0.150$^{\textrm{(a)}}$ \\
N & 1.0$^{\textrm{(a)}}$ & 0.8709 & 1.0$^{\textrm{(a)}}$ & 1.0$^{\textrm{(a)}}$
 \end{tabular}

{\medskip \centerline{\null\hspace*{2.5in}\footnotesize $^{\textrm{(a)}}$not varied}
\par}
 \end{table}%
\begin{table}[p] 
 \caption{Contributions to the standard deviation, $\chi^2$, from each data type for
each of the models. The number of data points contributing is in parentheses.  For each data type
the first row corresponds to the data set for which the model parameters were optimized, the
second row to the full data set.}
 \label{T2}
% \begin{tabular}{c|c|cccc} 
 \begin{tabular}{r|c|rrrr} 
Data & Data & & Models &  \\[-0.5ex]
type & set & GKex(01) & GKex(01-) & GKex(02L) & GKex(02S) \\
\hline\hline 
$G_{Mp}$ & opt & 43.3(68) & 43.6(68) & 48.1(68) & 47.9(68) \\
&full && same as above & \\
$G_{Ep}$ & opt & 67.2(48) & 48.2(44) &75.3(44) & 30.5(36) \\
& full & 67.2(48) & 74.8(48) & 112.2(48) & 136.8(48) \\
$G_{Mn}$ & opt &
122.4(35) & 120.2(35) & 121.0(35) & 122.7(35) \\
& full & & same as above & \\
$G_{En}$ & opt & 64.8(23) & 64.2(23) & 24.1(15) & 24.2(15) \\
& full & 65.3(24) & 65.0(24) & 68.2(24) & 68.3(24) \\ 
$R_p$ & opt & 29.0(17) & 22.6(17) & 23.1(21) & 11.8(21)l \\
& full & 114.0(21) & 106.5(21) & 23.1(21) & 11.8(21) \\
$R_n$ & opt & 0.0(0) & 0.0(0) & 0.6(3) &0.6(3) \\
& full & 9.6(3) & 17.7(3) &0.6(3) & 0.6(3) \\
\hline 
Total & opt & 326.7(191) & 298.9(187) & 336.3(195) & 237.7(178) \\
& full & 421.8(199) & 427.8(199) & 369.2(199) & 388.1 
\end{tabular}
\end{table} 

\section{Results}\label{s:4}
	Table~\ref{T1} presents the parameters  which minimize  $\chi^2$ for the above 4 cases.  For 
all 4 parameter sets the  hadronic form factor cut-off masses, $\Lambda_1$, 
$\Lambda_2$, $\Lambda_D$,  and $\mu_\phi$. are reasonable.  The relatively large value of
$\Lambda_2$, which controls the spin-flip suppression in QCD, is consistent with the slow approach
to asymptopia observed in polarized hadron scattering.  For the two cases in which $\Lambda_{QCD}$
is a fitted parameter, as well as the two for which it is fixed, it is consistent with high energy experiment.
The addition of the
$\omega'$(1.419) meson in GKex(02L) and GKex(02S) has moved $\kappa_\omega$ closer to the
expected small negative value than all earlier fits, but there is still the implication of some
effect
from a higher mass isoscalar meson.  The adequacy of the fits is an indication that the form factors
with more poles would be similar to those already obtained.

	In Table~\ref{T2} the values of $\chi^2$ are listed for the 4 cases and the 
contribution from each of the six form factor classes of measurement are detailed.  Also shown are
the values of $\chi^2$ when any data points omitted from the fit are re-inserted. 
\begin{figure}[hbt]
$$
\BoxedEPSF{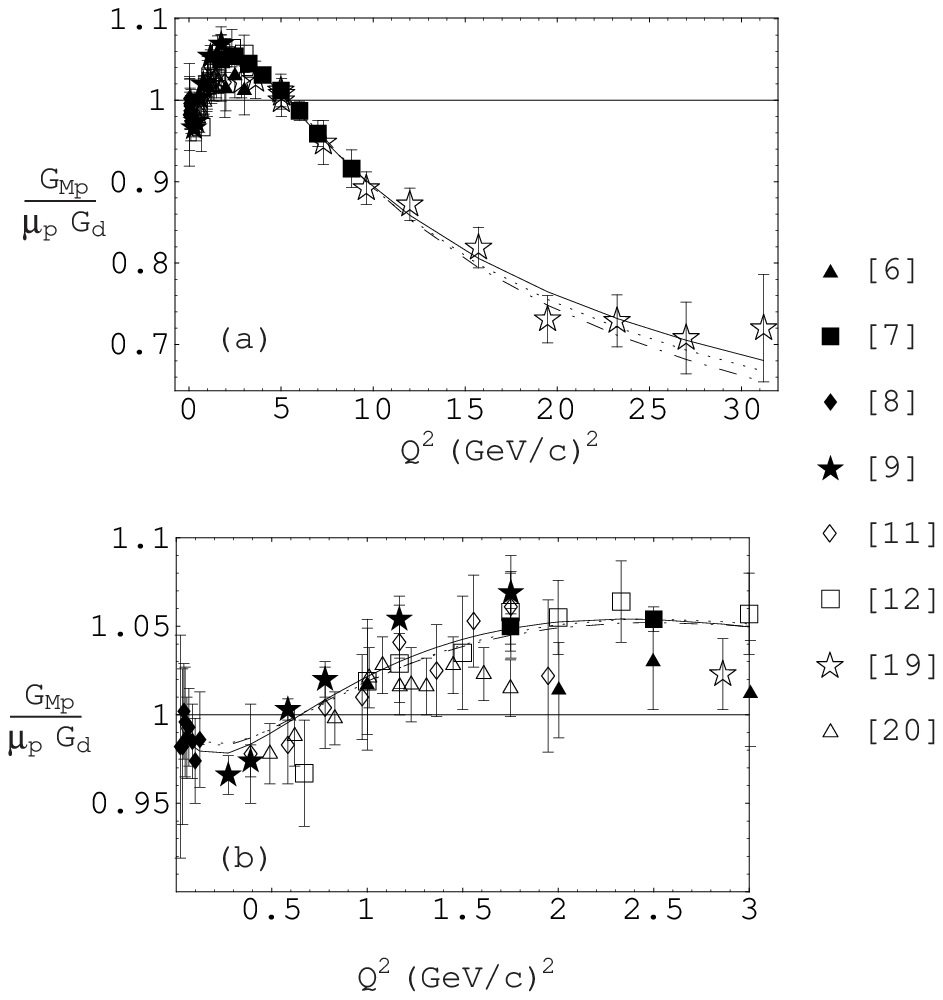 scaled 1200}
$$
\caption{$G_{Mp}$ normalized to $\mu_p G_d$.  Comparison of the models 
GKex(01) [solid], GKex(02L) [dotted] and GKex(02S ) [dash-dotted] with the data of \protect\Cite{RCW},
 \protect\Cite{LA}, \protect\Cite{FB}, \protect\Cite{KMH}, \protect\Cite{CHB}, \protect\Cite{WB73}, 
\cite{AFS} and \cite{PEB}.  (a) The full data range.  (b) Expansion of the range  
$Q^2 \leq 3.0$~GeV$^2/c^2$.
The data symbols are listed in the figure.} 
\label{elFig3}
\end{figure}

\begin{figure}[hbt] 
$$
\BoxedEPSF{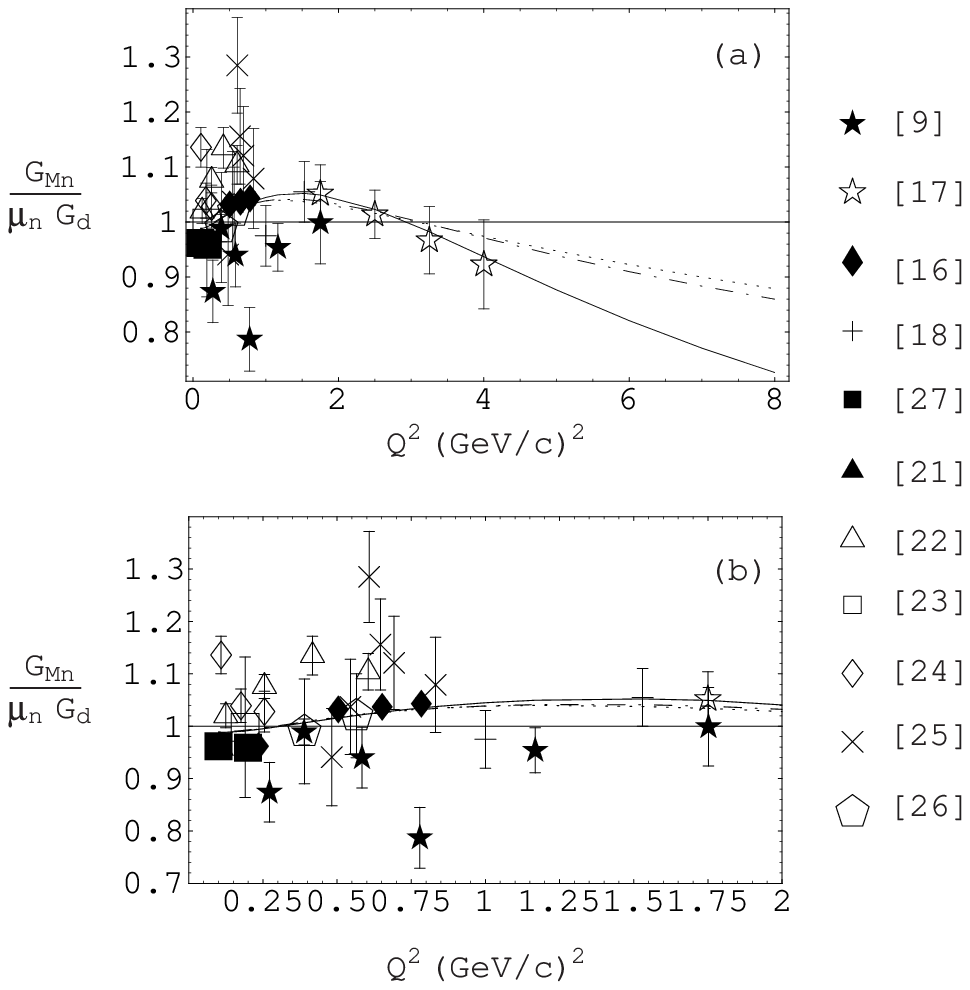 scaled 1200}
$$
\caption{$G_{Mn}$ normalized to $\mu_n G_{d}$.  Comparison of the models 
GKex(01) [solid], GKex(02L) [dotted] and GKex(02S ) [dash-dotted] with the data of \protect\Cite{KMH},  
 \protect\Cite{HA1}, \protect\Cite{AL}, \protect\Cite{WB72}, \Cite{HA2}, \Cite{Bruin}, \Cite{HG},
 \Cite{PM}, \Cite{ASE}, \Cite{WB69}, and \Cite{WX}.  
The data symbols are listed in the figure.  
 (a) The full data range.  (b) Expansion of the range  
$Q^2 \leq 2.0$~GeV$^2/c^2$.}
\label{elFig4}
\end{figure}
	We note, as can also be seen in Figs.~\ref{elFig3} and \ref{elFig4} that the quality of fit to
 the magnetic form factors, $G_{Mp}$ and $G_{Mn}$ changes negligibly as we refit to the datas sets that
differ in the electric form factors and the electric to magnetic form factor ratios.  As discussed in
\cite{EL}, the large excess of $\chi^2$ over the number of data points for $G_{Mn}$ is due to
obvious inconsistencies in the data set for $G_{Mn}$  at $Q^2 ~< ~0.8~\mbox{GeV}/c^2$.  The
displacement of adjacent data points well beyond their error bars in this range is evident in
the figures and contributes about 90 to the $\chi^2$ of $G_{Mn}$.

	The interesting changes are, of course, in the fits to $G_{Ep}$, $G_{En}$, $R_p$ and $R_n$.  As
noted in the introduction, removing the 4 very high values of  $G_{Ep}$ data \cite{RCW} does
surprisingly little to allow a better fit to the $R_p$ data already in GKex(01).  Several of the
parameters, all three $\kappa_m$, $\Lambda_2$ and $\Lambda_D$, have large changes (see
Table~\ref{T1}), but this results in a small shift between the predictions of GKex(01 and GKex(01-)
as is evident in Table~\ref{T2} and Figs.~\ref{elFig1} and \ref{elFig2}.  The figures show a slightly
better fit to $R_p$ correlated with a very slightly worse fit to $G_{Ep}$.  The former is reflected in
Table~\ref{T2} by the decrease  in the $\chi^2$ contribution of the 17  $R_p$ points to which
those cases were optimized from 29.0 t0 22.6.  When the 4 higher $Q^2$ of \cite{CP2} are
added the $\chi^2$ contribution is much larger than the number of points (21).  The drop in the
$\chi^2$ contribution to $G_{Ep}$ from 67.2 to 48.2 is entirely due to the omission of the 4 data points
of \cite{RCW}, but the $\chi^2$ for the full set of 48 points is a little larger because of the
competition with $R_p$.  The implication is that there is a constraint on the fit to $R_p$ from data
independent of $G_{Ep}$, arising from the model correlations between all the nucleon emff.  This is
shown to be the case below.

\begin{figure}[hbt]
$$
\BoxedEPSF{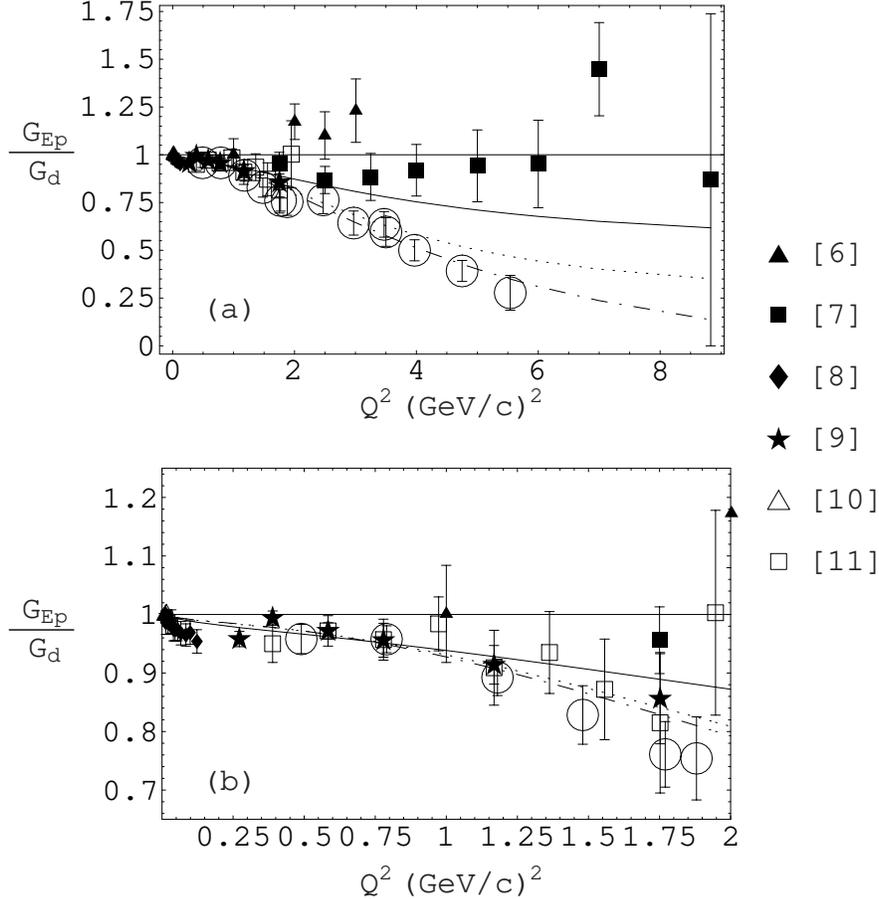 scaled 1200}
$$
\caption{$G_{Ep}$ normalized to $G_d$.  Comparison of the models 
GKex(01) [solid], GKex(02L) [dotted] and GKex(02S ) [dash-dotted] with the data. 
The data references are the same as in Fig. 1 and the data symbols are listed in the figure.
The points labelled by open
circles are obtained by multiplying $R_p$ data \protect\Cite{CP1} and \protect\Cite{CP2} by 
the $G_{Mp}$ of GKex(02S)
normalized by $\mu_p G_d$.  (a) The full data range.  (b) Expansion of the range
$Q^2 \leq 2.0$~GeV$^2/c^2$.}
\label{elFig5}
\end{figure}

\begin{figure}[hbt] 
$$
\BoxedEPSF{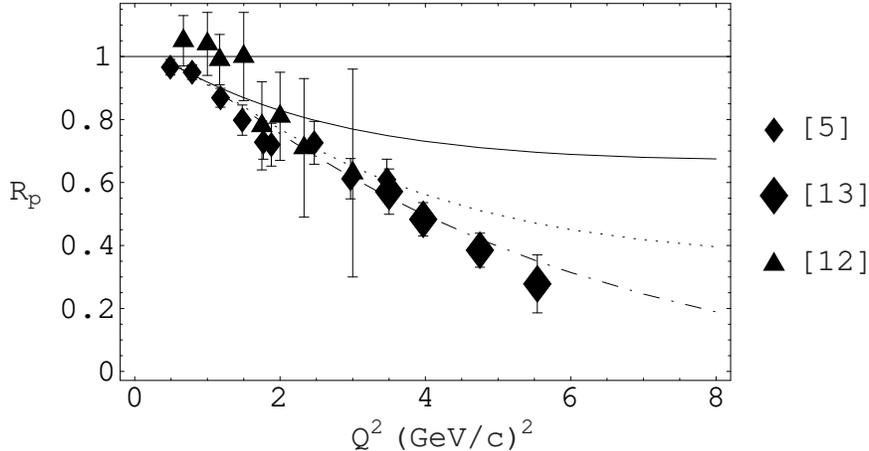 scaled 1200}
$$
\caption{$R_p$, the ratio $\mu_p G_{Ep}/G_{Mp}$.  Comparison of the models 
GKex(01) [solid], GKex(02L) [dotted] and GKex(02S ) [dash-dotted] with the data.  
The data is from \protect\Cite{CP1}, \protect\Cite{WB73}, and \protect\Cite{CP2}.}
\label{elFig6}
\end{figure}

\begin{figure}[hbt] 
$$
\BoxedEPSF{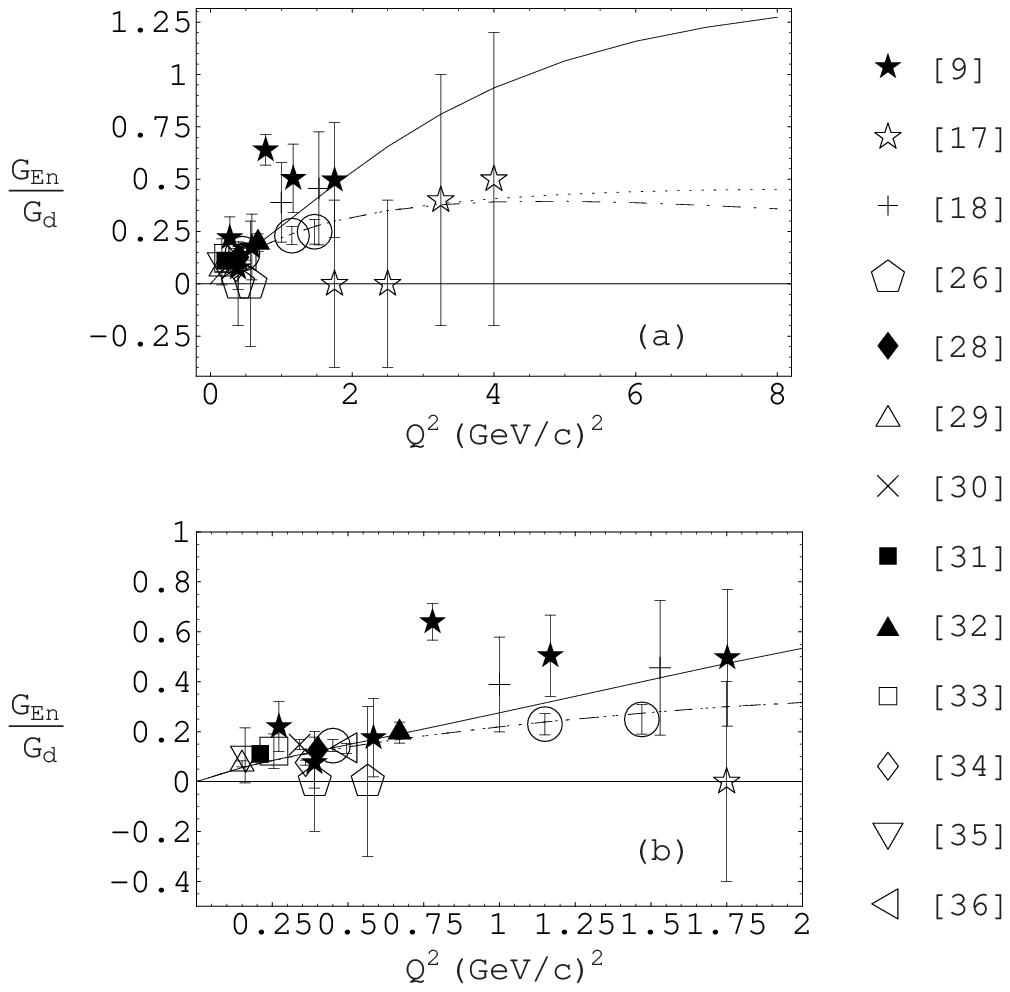 scaled 1200}
$$
\caption{$G_{En}$ normalized to $G_{d}$.  Comparison of the models 
GKex(01) [solid], GKex(02L) [dotted] and GKex(02S ) [dash-dotted] with the data  of
\protect\Cite{KMH}, \protect\Cite{AL}, \protect\Cite{WB72}, \protect\Cite{WB69},
 \protect\cite{JG}, \protect\cite{CH}, \protect\cite{MO}, \protect\cite{IP}, \protect\cite{DR},
 \protect\cite{TE}, \protect\cite{MM}, \protect\cite{J-W} and \protect\cite{HZ}.
The data of \protect\cite{MO}, \protect\cite{DR} and \protect\cite{MM} are the reevaluated values of
\protect\cite{TW}.  The slope  at $Q^2=0$ is from \protect\cite{SK}.
The points labelled by open
circles are obtained by multiplying $R_n$ data \protect\Cite{RM} by the $G_{Mn}$ of GKex(02S)
normalized by $\mu_n G_d$.  (a)  $Q^2 \leq 8.0$~GeV$^2/c^2$.  (b) Expansion of the range $Q^2 \leq 
2.0$~GeV$^2/c^2$.}
\label{elFig7}
\end{figure}

\begin{figure}[hbt]
$$
\BoxedEPSF{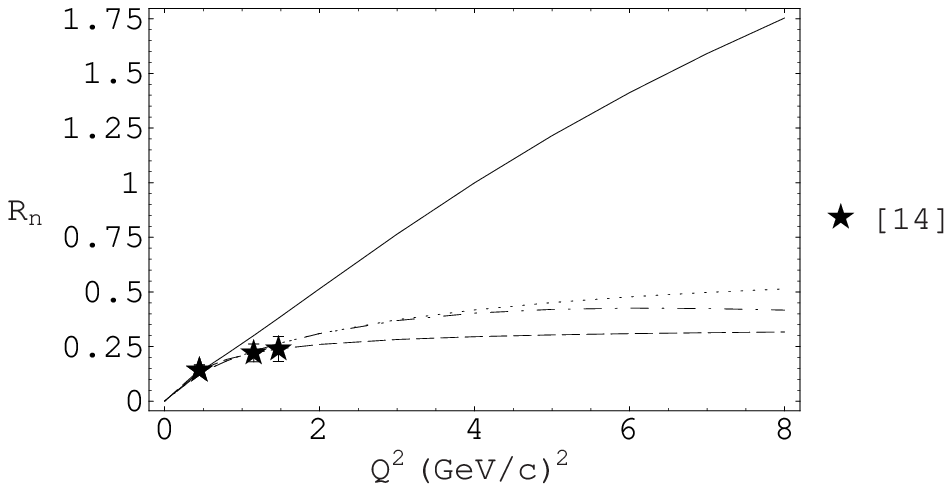 scaled 1200}
$$
\caption{$R_n$, the ratio $\mu_n G_{En}/G_{Mn}$.  Comparison of the models 
GKex(01) [solid], GKex(02L) [dotted] and GKex(02S ) [dash-dotted] with the data. 
The dashed curve is  $R^{Galster}_n(Q^2)$ of Eq.~(\ref{e1}).  The ``experimental" points are
described in the text \protect\Cite{RM}.}
\label{elFig8}
\end{figure}

	Substituting the new $R_n$ values for the conflicting $G_{En}$ data of \cite{KMH},  \cite{AL}
and \cite{WB72} causes a large difference between the GKex(02L) and GKex(01-) fits to $G_{Ep}$,
$G_{En}$, $R_p$ and $R_n$, as seen in Table~\ref{T2} and Fig.~\ref{elFig5} - Fig.~\ref{elFig8}. 
In particular  for GKex(02L) the $\chi^2$ contribution for all 21 $R_p$
data points is 23.1.  Fig.~\ref{elFig6} shows the strong improvement in the fit to $R_p$. 
 The figure also
shows that the goodness of the $\chi^2$ value is somewhat misleading because that fit is
systematically high for the 3 highest $Q^2$ data points.  On the other hand the $\chi^2$
contribution for all 44 $G_{Ep}$ data points increases from 48.2 in GKex(01-) to 75.3 in GKex(02L)
because of the compromise of better fitting $R_p$.  The $\chi^2$ contribution for the 3
$R_n$ points now included is only 0.6.  $G_{En}$ now contributes 24.1 for the remaining 15 data
points (which still include highly scattered low $Q^2$ data as discussed in \cite{EL}) instead of 64.8
for the 23 data points in GKex(01-).

	For the GKex(02S) case the $G_{Ep}$ data of \cite{LA}, which is clearly inconsistent with the new
$R_p$ data \cite{CP1} and \cite{CP2}, is also omitted.  The results are very good if the modern data
is chosen when in conflict with the older Rosenbluth separation results.  The $\chi^2$ contribution
from the remaining 36 $G_{Ep}$ points is only 30.6 and for all 21 $R_p$ points only11.8.  For the
remaining types of form factor measurements there is a negligible change of $\chi^2$ between the
GKex(02L) and GKex(02S) cases.

Figs.~\ref{elFig5}, \ref{elFig6}, \ref{elFig7},  and \ref{elFig8} show the successive
improvements in $G_{Ep}$,
 $R_p$, $G_{En}$, and $R_n$ as the optimization data sets are varied from GKex(01) to GKex(02L) and
to GKex(02S).  To demonsrate the  correlation between the electric form factors and the ratio of
electric to magnetic form factors we have, in Figs.~\ref{elFig5} and \ref{elFig7}, entered 
(as circles) the
electric form factor values obtained by multiplying the experimental $R_p$ and $R_n$  values by
the case GKex(02S) model values of the magnetic form factors normalized by the magnetic
moments.  The correlation with the model prediction for the electric form factors is excellent. 

	The figures show the  model extrapolations of $R_p$, $G_{Mn}$, $G_{En}$ and $R_n$ up to
$Q^2$ of 8 ~GeV$^2/c^2$ for the guidance of future experiments and because of their relevance to deuteron and
other nuclear electromagnetic scattering predictions.  The extrapolation is sensitive to 
the weight given to the polarized
vs. the Rosenbluth separation data in the fits.  The resolution of this dichotomy will, in the context of the
physical model employed here, greatly restrict the nucleon emff over a large range of momentum transfer.

	The polarization measurements of $R_p$ and $R_n$ may soon be extended to larger $Q^2$, so it is
of interest to examine the predictions of the good fit GKex(02S) as $Q^2$ increases.  As seen in 
Fig.~\ref{elFig5} the model curve is, as is the data, approximately linear in the range  $0.5
$~GeV$^2/c^2 \leq Q^2 \leq 5.6$~GeV$^2/c^2$, but the model curve's slope is gradually decreasing in 
manitude.  A linear fit to the data would change sign near $Q^2 = 8$ ~GeV$^2/c^2$ where the model 
predicts 0.19.  The model crosses zero near $Q^2 = 14$ ~GeV$^2/c^2$ with a very small slope.

	Fig.~\ref{elFig8} shows the Galster curve, $R^{Galster}_n(Q^2)$ of Eq.~(\ref{e1}), to compare with
the model and the data.  The model fits the data but deviates from the Galster curve after that.  The
model increases faster, reaching a maximum of 0.426 at $Q^2 = 4$ ~GeV$^2/c^2$
 where the Galster value is only 0.309, while $R^{Galster}_n(Q^2)$
increases monotonically to an asymptotic value of 0.342.  A measurement of the present quality at
 $Q^2 = 4$ ~GeV$^2/c^2$ could distinguish between the two.

\section{Conclusions}\label{s:5}
		The GKex model, consistent with vector meson dominance and perturbative
QCD in the appropriate momentum transfer regions, represents well a consistent
set of neutron and proton emff.  This set includes polarization measurements,
which are directly related to the ratios of electric to magnetic form factors, and
differential cross section measurements of the magnetic form factors.  The values
of the electric form factors from the Rosenbluth separation of the differential
cross section is,  in our final selection GKex(02S), only used for the lower range of
$Q^2$ where the magnetic contributions are less dominant.  Because of the
physical properties of the model and the good quality of the fit we expect that
the model predictions are sufficiently accurate to be used for predictions of the
electromagnetic properties of nuclei.  The model values may also be useful in
planning future experiments.

	The above conclusions are only valid to the extent that adequate physics is included in the
GKex models.  Only the $\rho$ meson exchange includes the width of the vector mesons 
(from dispersion relation results).  There will be corrections from the widths of the other exchanged 
vector mesons.  However the next most important, the $\omega$ and $\phi$, have very narrow widths.  The 
higher masses of the $\rho'$ (1450) and the $\omega'$ (1420) reduces the importance of their substantial
widths because of their distance from the physical region and their partial replacement by the pQCD
term.

	In assuming vector dominance we have neglected the multi-meson exchange continuum contributions.
In particular the two-pion continuum may have an influence at very low $Q^2 \leq 0.4$~GeV$^2/c^2$.
Indeed, as remarked in\cite{EL} and can be seen in Fig. 5b, the $G_{Ep}$ data of \cite{FB} 
has a more negative
 slope for $Q^2 \leq 0.3$~GeV$^2/c^2$ than the higher $Q^2$ data and the model fit.  The addition of a 
2-pion exhange term to the model may enable a change of slope between the two regions, but would have 
little effect on the model fit for $Q^2 \geq 0.5$~GeV$^2/c^2$.

	One may also want to consider some higher mass vector mesons.  This would have some importance
in the fits of \cite{MMD} and\cite{HO}, but are much less important in these GK type models because 
of the transition to pQCD behavior.

\subsection*{Acknowledgments}
	The author is grateful to Charles Perdrisat and Richard Madey for timely
information about their polarization measurements.

\end{document}